\documentclass[letter,aps,prd,10.5pt,preprintnumbers,showpacs,showkeys,superscriptaddress,nofootinbib,amsmath,amssymb,floatfix
,twocolumn
]{revtex4-2}

\usepackage{times}

\usepackage[hidelinks]{hyperref}
\usepackage{amsmath}
\usepackage{amssymb}
\usepackage{graphicx}

\usepackage{subfigure}
\usepackage{booktabs}
\usepackage{rotating}

\usepackage{bm}
\usepackage{float}
\usepackage{epstopdf}

\usepackage{xcolor}
\usepackage{comment}
\usepackage{graphicx}
\usepackage{subfigure}
\usepackage[T1]{fontenc}
\usepackage{lmodern,microtype}
\usepackage{amsmath,amssymb,bm,mathtools}
\usepackage{graphicx,booktabs,array,tabularx}
\usepackage[dvipsnames]{xcolor}
\usepackage{caption,subcaption}
\usepackage{enumitem}
\usepackage{fancyhdr}

\definecolor{KRblue}{RGB}{22,63,103}
\definecolor{KRred}{RGB}{157,45,38}
\setlist{nosep,leftmargin=*}
\numberwithin{equation}{section}
\newcommand{\dd}{\mathrm d}
\newcommand{\eff}{\mathrm{eff}}
\newcommand{\KR}{\mathrm{KR}}

\makeatletter

\newcommand{\Rmnum}[1]{\expandafter\@slowromancap\romannumeral #1@}

\makeatother

\makeatletter
\newcommand*{\rom}[1]{\expandafter\@slowromancap\romannumeral #1@}
\makeatother

\begin{document}

\title{Causal structure and optical signatures of rotating power law Kalb-Ramond geometry}

\author{Hassan  Hassanabadi}
\email{hassanhassanabadi@mail.fresnostate.edu}
\affiliation{Department of Physics, California State University, Fresno,
	Fresno, California 93740, USA.}

\author{Orlando Luongo}
\email{orlando.luongo@unicam.it}
\affiliation{Universit\`a di Camerino, Via Madonna delle Carceri, Camerino, 62032, Italy.}
\affiliation{INAF - Osservatorio Astronomico di Brera, Milano, Italy.}
\affiliation{Istituto Nazionale di Fisica Nucleare, Sezione di Perugia, Perugia, 06123, Italy.}
\affiliation{SUNY Polytechnic Institute, 13502 Utica, New York, USA.}
\affiliation{Al-Farabi Kazakh National University, Al-Farabi av. 71, 050040 Almaty, Kazakhstan.}

\begin{abstract}
We investigate a stationary and axisymmetric power law Kalb-Ramond geometry,
obtained as a rotating extension of the asymptotically flat static solution. We reconstruct
the exact inverse Einstein source supporting the geometry and determine its
energy conditions and curvature properties. We characterize the horizon and stationary limit configurations and quantify the
outer ergoregion through its angular dependent radial thickness plus its
coordinate and proper spatial volumes. We derive the unstable spherical
photon region and show that the Kalb-Ramond deformation shifts the prograde
photon orbit into the ergoregion at lower spin, while the retrograde region
remains outside throughout the entire domain. From the
unstable photon region, we construct the exact vacuum critical curve and
analyze its area equivalent diameter and displacement, including
comparisons with M87$^\ast$ and Sgr~A$^\ast$ within the asymptotically
admissible domain. We further examine the fixed-$J$ geometric behavior and
rotational energy extraction, with emphasis to rotating weak deflection and two dimensional
weak lensing, commenting the physical consequences of our solution in terms of observable outcomes.
\end{abstract}



\maketitle

\section{Introduction}

The Kalb-Ramond (KR) field was introduced by Kalb and Ramond as an
antisymmetric rank-two potential $B_{\mu\nu}$ coupled  to a closed
string world sheet, in  analogy to the coupling of a vector potential
to a point particle \cite{KalbRamond1974}. Afterwards, it became part of the
massless bosonic sector of heterotic string theory \cite{Gross1985}. More precisely, its
gauge invariant field strength is the three-form
$H_{\mu\nu\rho}=\partial_{[\mu}B_{\nu\rho]}$, which in four dimensions may
be dualized locally to a pseudoscalar. Accordingly, these properties make the KR field a
natural bridge between strings and extended theories of gravity.

A particularly useful realization occurs when $B_{\mu\nu}$ develops a
nonzero vacuum expectation value and couples nonminimally to curvature. Accordingly, the
vacuum selects preferred tensor directions and can spontaneously break
local Lorentz symmetry while preserving coordinate covariance
\cite{Altschul2010}.  The associated three-form can also be interpreted as a
source of effective spacetime torsion \cite{Majumdar1999}, with torsional
line defect geometries providing a related gravitational puzzle
\cite{Letelier1995}. In this respect, additional
KR phenomenology includes parity sensitive cosmic microwave background signals
\cite{Maity2004}, tensor field cosmology and inflation \cite{Li2012,Aashish2019},
wormhole stability \cite{Cox2016}, strong lensing in extra-dimensional
settings \cite{Chakraborty2017}, Solar System lensing and perihelion tests
\cite{Kar2003}, quantum consistency of Lorentz-violating antisymmetric
tensors \cite{AashishPanda2019} and so on\footnote{String-inspired environmental violations of
CPT symmetry provide another application of antisymmetric-background physics
\cite{Ellis2013}.  The smallness of weak field corrections does
not preclude appreciable effects near compact objects, where causal surfaces,
null orbits, thermodynamics and rotational energy extraction probe the
geometry nonlinearly.}.

The exact solution obtained by Lessa \emph{et al.} \cite{Lessa2020} provides
the static seed used throughout this manuscript. In this case, starting from a
self-interacting KR field with a nonminimal curvature coupling, the authors a spherically-symmetric metric function of the form
$F(r)=1-2M/r+\Upsilon/r^{2/\lambda}$, where the additional term changes the
horizon structure and, for $\lambda\rightarrow1$, reduces to a
Reissner-Nordstr\"om spacetime\footnote{The correction vanishes at infinity, leading to a Minkowski spacetime while exhibiting a nonzero tensor
background.  This makes the solution useful for separating strong field KR
signatures from a cosmological asymptotic contribution. However, KR black holes do not possess a universal asymptotic structure since the outcome depends on the action, vacuum configuration, matter sector, etc.}. Other static solutions generated
by alternative KR vacuum ansatz turn out to be  Schwarzschild-like, showing a
radially rescaled or solid angle deficit geometry, departing from the standard
Minkowski space \cite{Yang2023}.  Electrically charged KR solutions exist
both with and without a cosmological constant \cite{Duan2024}.

KR
backgrounds have also been investigated in higher-form branes
\cite{Fu2012}, cosmological bounces and anisotropic cosmology
\cite{Nair2022,MalufNeves2022} and traversable wormholes
\cite{LessaWormhole2021,MalufMuniz2022}.  Global monopole studies,  exhibiting solid angle
deficits and asymptotics \cite{Rhie1991}, provide a useful
comparison class for lensing in Lorentz violating geometries
\cite{Ovgun2018,Kuang2022}.  More recently, monopoles have been coupled
directly to Ricci-coupled KR bumblebee gravity \cite{Belchior2025}, while
black hole shadows and lensing have been examined in Lorentz-violating
monopole backgrounds \cite{Ovgun2025,Pantig2025}.

Astrophysical applications are motivated by the fact that the same parameters
that characterize the KR vacuum enter observable strong and weak field
quantities.  Static and rotating analyses have addressed particle dynamics,
weak lensing and energy processes \cite{Atamurotov2022}, fermionic greybody
factors and quasinormal modes \cite{AlBadawi2023,Baruah2023} and parameter
constraints from shadows, quasinormal frequencies and other optical
properties \cite{Zubair2023,Filho2024}.  KR backgrounds have also been
studied in traversable-wormhole lensing \cite{Sarkar2025}, nonsingular
cosmological evolution \cite{Nair2022}, parity violating gravity
\cite{Manton2024} and cosmological quantum entanglement \cite{Paul2020}.

Since according to the Kerr hypothesis and to thermodynamic considerations, astrophysical black holes are expected to rotate, extending the static
seed appears essential, i.e., frame dragging splits prograde and retrograde photon
orbits, creates an ergoregion, enables Penrose extraction and superradiance
and, then, introduces degeneracies between spin and the KR deformation in shadow
observables.

In this respect, Kumar, Ghosh and Wang \cite{Kumar2020} constructed the axisymmetric
power law KR geometry and determined its horizons and stationary-limit
surfaces, Komar mass and angular momentum, generalized Smarr relation, photon
capture region, shadow observables, M87$^*$ constraints and finite-distance
weak deflection corrections\footnote{In weak gravity regime, optical effects may be related to thermodynamics and appear well-established, see Ref. \cite{luongo2025}. Thus, departing from them would falsify non-minimally coupled general relativity, even at the level of low energies.}.

Motivated by the above results, we explore a the rotating
power law KR geometry at the level of its effective source, causal
structure, null dynamics and observable signatures. We reconstruct the exact
inverse Einstein source associated with the rotating metric and determine its
principal stresses, energy conditions, curvature properties and relevant
limiting geometries. Afterwards, we analyze the horizon and stationary limit
surfaces, introducing quantitative measures of the ergoregion through its
angular thickness and both coordinate and proper volumes. The null sector is
treated by deriving the spherical photon-orbit conditions, separating the
prograde and retrograde regions and defining the corresponding
photon ergosurface gaps. This allows us to determine how the KR
deformation shifts the spin at which the prograde photon orbit penetrates the
ergoregion. From the unstable spherical photon region, we construct the exact
vacuum critical curve, together with its area equivalent diameter and
horizontal displacement. In this respect, we adopt these quantities for direct comparisons
with the M87$^\ast$ and Sgr~A$^\ast$ observations within the asymptotically
admissible domain $0<\lambda<2$.

In addition, we investigate the consistency of the surface gravity description
with a fixed-$J$ first law, leading to geometric response functions whose
thermodynamic interpretation remains conditional on the entropy and conserved
charges of the underlying rotating KR theory. Further, we  study
rotational energy extraction and derive the rotating weak deflection expansion. In particular, we formulate weak lensing through a two dimensional lens map and its full
Jacobian, from which the signed and unresolved magnifications are thus computed\footnote{Rotating intensity and polarization maps for thick accretion flows in the same
power law geometry have also been investigated by Yang, Ye and Zeng
\cite{YangYeZeng2026}. Those observables depend on the emission model and
accretion dynamics, while the quantities considered here are determined by
the spacetime geometry and null propagation. In particular, the shadow
observable used throughout this work is the vacuum null critical curve defined
by the unstable photon region and is therefore distinct from a
model-dependent brightness ring.}.

The paper is organized as follows.  Section 2 introduces the static seed and
rotating geometry, reconstructs its effective source and states the precise
field-equation status.  Section 3 analyzes horizons and the ergoregion.
Sections 4 and 5 develop null geodesics, photon orbits, shadows and
observational deviations.  Sections 6-9 discuss surface gravity and
geometric response functions,
rotational-energy extraction, rotating weak deflection and magnification,
followed by  conclusions and perspectives in view of future works.

\section{Rotating geometry, effective source and dynamical status}

The term ``KB black hole'' identifies the antisymmetric-field sector, identifying a class of solutions.  String effective rotating solutions and the power law KB branch introduced by Lessa \emph{et al.} approach Minkowski spacetime, with the standard ADM interpretation being cleanest when the deformation decays at least as rapidly as the mass term \cite{Lessa2020,Kumar2020}. Other exact Lorentz-violating KR backgrounds yield Schwarzschild- or Reissner-Nordstr\"om-like geometries with a constant asymptotic normalization. These spacetimes can be locally flat at large radius while retaining a solid-angle deficit or excess \cite{Yang2023,Duan2024}. When a cosmological constant is present, the corresponding branches are asymptotically de Sitter or anti-de Sitter and a global-monopole sector can introduce an additional angular deficit \cite{Yang2023,Duan2024,Belchior2025}.

These differences should not be regarded as inconsistencies among the cited studies. They arise from distinct actions, nonminimal couplings, vacuum configurations of the antisymmetric tensor, matter sectors and integration branches. The present analysis deliberately adopts the asymptotically Minkowskian power law family and focuses on the parameter range in which the deformation falls at least as fast as the mass contribution. This choice permits the conventional normalization of mass, impact parameter and distant-observer shadow observables, while the deficit-angle and cosmological branches remain valuable complementary realizations of KR gravity.

\subsection{Static seed and rotating construction}

The static power law seed obtained in Ref.~\cite{Lessa2020} is
\begin{equation}
\begin{aligned}
 \dd s^2&=-F(r)\dd t^2+\frac{\dd r^2}{F(r)}+r^2\dd\Omega^2,\\
 F(r)&=1-\frac{2M}{r}+\frac{\Upsilon}{r^{2/\lambda}} .
\end{aligned}
 \label{eq:seed}
\end{equation}
It is convenient to introduce
\begin{equation}
 q=2-\frac{2}{\lambda},\qquad
 m(r)=M-\frac{\Upsilon}{2}r^{1-2/\lambda},
 \label{eq:defq}
\end{equation}
so that $F=1-2m(r)/r$.  Applying the non-complexification rotating
construction developed by Azreg-A\"{\i}nou \cite{Azreg2014} gives
\begin{equation}
\begin{split}
 \dd s^2={}&-\frac{\Delta}{\Sigma}(\dd t-a\sin^2\theta\,\dd\phi)^2
 +\frac{\Sigma}{\Delta}\dd r^2+\Sigma\dd\theta^2\\
 &+\frac{\sin^2\theta}{\Sigma}
 [(r^2+a^2)\dd\phi-a\dd t]^2,
 \label{eq:metric}
\end{split}
\end{equation}
where $
 \Sigma=r^2+a^2\cos^2\theta,\qquad
 \Delta=r^2+a^2-2Mr+\Upsilon r^q$.

This is algebraically the rotating metric of Kumar, Ghosh and Wang after identifying their parameters $s=\lambda$ and $\Gamma=\Upsilon$ \cite{Kumar2020}.  The equality represents only a geometric relation and, so, it cannot be seen as a proof of the complete rotating field equations.

We use the dimensionless variables $
 x=\frac{r}{M}$, $a_*=\frac{a}{M}$, $
 \gamma=\frac{\Upsilon}{M^{2/\lambda}}$. For $0<\lambda<2$ the KB deformation term falls faster than $1/r$, providing the cleanest conventional ADM interpretation.  At $\lambda=2$ it is absorbed by $M\mapsto M-\Upsilon/2$. For $\lambda>2$ it falls more slowly than the mass term and standard asymptotic charges require additional care.

\subsection{Inverse Einstein source}

We define the source required by the geometry through
\begin{equation}
 T^{\eff}_{\mu\nu}=\frac{1}{\kappa}G_{\mu\nu}[g],
 \qquad \kappa=8\pi G.
 \label{eq:inverse}
\end{equation}
With the orthonormal coframe
\begin{align}
 \omega^{\hat0}&=\sqrt{\frac{\Delta}{\Sigma}}
 (\dd t-a\sin^2\theta\dd\phi),\nonumber\\
 \omega^{\hat1}&=\sqrt{\frac{\Sigma}{\Delta}}\dd r,
 \qquad \omega^{\hat2}=\sqrt{\Sigma}\dd\theta,\nonumber\\
 \omega^{\hat3}&=\frac{\sin\theta}{\sqrt\Sigma}
 [(r^2+a^2)\dd\phi-a\dd t].
\end{align}
The source has eigenvalues
\begin{equation}
 T^{\hat\mu}{}_{\hat\nu}=\mathrm{diag}(-\rho,p_r,p_\perp,p_\perp),
 \qquad p_r=-\rho,
 \label{eq:eigen}
\end{equation}
with the exact expressions
\begin{align}
 \kappa\rho&=\frac{2r^2m'}{\Sigma^2}
 =\frac{\Upsilon(2/\lambda-1)r^{2-2/\lambda}}{\Sigma^2},
 \label{eq:rho}\\
 \kappa p_\perp&=-\frac{r\Sigma m''+2a^2\cos^2\theta\,m'}{\Sigma^2}\nonumber\\
 &=\frac{\Upsilon(2/\lambda-1)r^{-2/\lambda}}{\Sigma^2}
 \left[\frac{r^2}{\lambda}+\left(\frac1\lambda-1\right)a^2\cos^2\theta\right].
 \label{eq:pperp}
\end{align}
These quantities are finite at every nonzero regular horizon. So, apparent divergences of $T_{rr}=p_r\Sigma/\Delta$ are due to the choice of Boyer-Lindquist coordinates only.

\subsection{Curvature structure and limiting geometries}

Several checks are immediate.  First, $\Upsilon=0$ gives $T^{\eff}_{\mu\nu}=0$ and the Kerr metric.  Second, $a=0$ gives
\begin{equation}
 \begin{aligned}
 \kappa\rho
 &=\Upsilon\left(\frac{2}{\lambda}-1\right)
 r^{-2-2/\lambda},\\[2pt]
 p_r&=-\rho,
 \qquad p_\perp=\frac{\rho}{\lambda}.
 \end{aligned}
\end{equation}
Third, the Ricci scalar reconstructed from the trace is
\begin{equation}
 \begin{aligned}
 R
 &=\frac{2\bigl(2m'(r)+r m''(r)\bigr)}{\Sigma}\\[2pt]
 &=\frac{2\Upsilon}{\Sigma}
 \left(\frac{2}{\lambda}-1\right)
 \left(1-\frac{1}{\lambda}\right)r^{-2/\lambda}.
 \end{aligned}
 \label{eq:Ricci}
\end{equation}
It vanishes for Kerr, for the traceless $\lambda=1$ Kerr-Newman-like limit and for the shifted-mass $\lambda=2$ limit.  Finally, $
 \nabla_\mu T_{\eff}^{\mu\nu}=0$ holds identically by the contracted Bianchi identity.

The vanishing of the Ricci scalar in these special limits does not imply
regularity.  Indeed, for generic $\lambda\ne1,2$, nonzero $a$ and fixed
$\theta\ne\pi/2$, Eq. \eqref{eq:Ricci} behaves as
$R\sim r^{-2/\lambda}/(a^2\cos^2\theta)$ as $r\to0$. On the equatorial plane
the divergence is stronger, $R\sim r^{-2-2/\lambda}$.  Thus the generic
$r=0$ singular set is not confined to the Kerr ring.  The static limit also
has a singular center with the latter radial scaling.  At the exceptional
values $\lambda=1$ and $\lambda=2$ the Ricci scalar vanishes, but quadratic
invariants retain the familiar Kerr-Newman-like or Kerr ring singularity,
respectively.  The limiting geometries therefore have distinct
interpretations, i.e., $\Upsilon=0$ gives Kerr, $a=0$ returns the exact static KR
seed, $\lambda=1$ is geometrically Kerr-Newman-like without requiring an
electromagnetic charge and $\lambda=2$ shifts the coefficient of the
$1/r$ mass term.  These limits independently check the horizon, photon-orbit,
and thermodynamic expressions derived below.

For $\Upsilon>0$, $\rho\ge0$ for $0<\lambda\le2$.  The radial null energy condition is saturated.  The transverse condition reads
\begin{equation}
\begin{split}
 \kappa(\rho+p_\perp)={}&\frac{\Upsilon(2/\lambda-1)r^{-2/\lambda}}{\Sigma^2}\\
 &\times\left[\left(1+\frac1\lambda\right)r^2
 +\left(\frac1\lambda-1\right)a^2\cos^2\theta\right].
\end{split}
\end{equation}
For the parameter ranges plotted below it is satisfied outside $r_+$.  The transverse dominant condition follows from
\begin{equation}
 \frac{p_\perp}{\rho}=\frac1\lambda+\left(\frac1\lambda-1\right)
 \frac{a^2\cos^2\theta}{r^2}.
\end{equation}
It is violated for $\lambda=0.5$, saturated at $\lambda=1$ and satisfied outside the event horizon for the displayed $\lambda=1.5$ examples.

\subsection{Coordinate components and physical interpretation}

Although Eq. \eqref{eq:eigen} is diagonal in the locally orthonormal frame, rotation
produces a nonzero $t\phi$ component in Boyer-Lindquist coordinates.  The
nonvanishing covariant components are
\begin{align}
 T_{tt}&=\frac{\rho\Delta+p_\perp a^2\sin^2\theta}{\Sigma},\nonumber\\
 T_{t\phi}&=-\frac{a\sin^2\theta}{\Sigma}
 \left[\rho\Delta+p_\perp(r^2+a^2)\right],\nonumber\\
 T_{\phi\phi}&=\frac{\sin^2\theta}{\Sigma}
 \left[\rho a^2\Delta\sin^2\theta
 +p_\perp(r^2+a^2)^2\right],\label{eq:Tcoordinate}\\
 T_{rr}&=p_r\frac{\Sigma}{\Delta},\qquad
 T_{\theta\theta}=p_\perp\Sigma.\nonumber
\end{align}
The off-diagonal component describes rotational momentum transport in the
coordinate basis.  The local
source remains the anisotropic type-I tensor in Eq. \eqref{eq:eigen}.

\begin{figure*}[t]
 \centering
 \includegraphics[width=1.\textwidth]{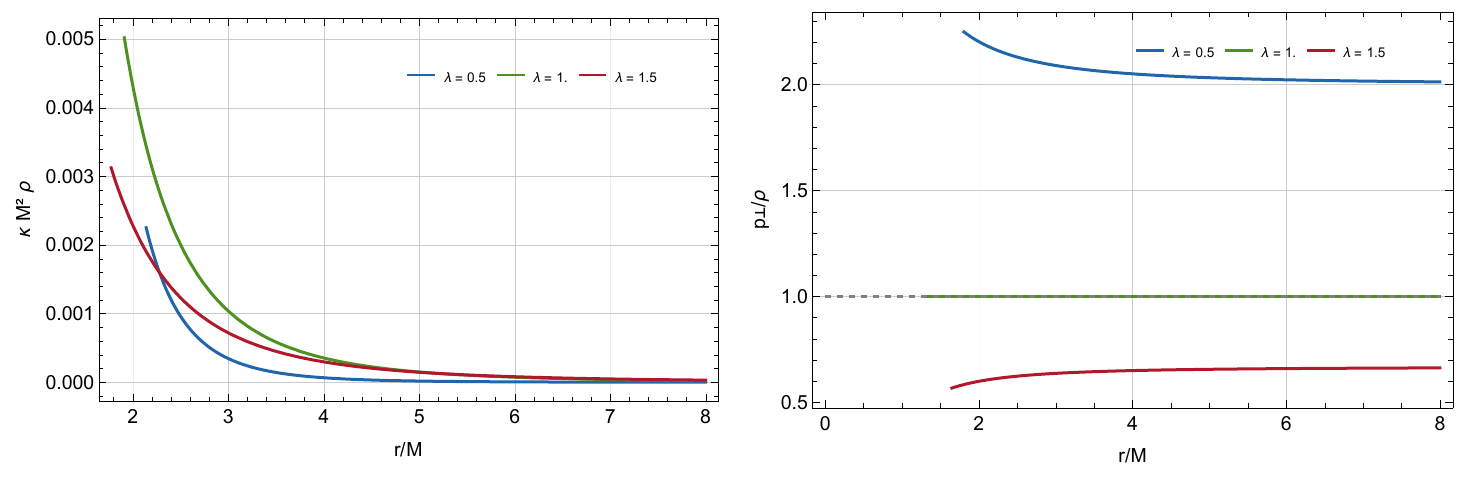}
 \caption{\emph{Inverse Einstein source outside the event horizon.  The left panel
 shows the energy density along the rotation axis.  The right panel makes the
 transverse dominant energy test evident, i.e.,  $p_\perp/\rho>1$ for
 $\lambda=0.5$, equality is reached for $\lambda=1$ and the displayed
 $\lambda=1.5$ branch lies below the DEC boundary.}}
 \label{fig:source}
\end{figure*}

Because $p_r=-\rho$, the radial null-energy condition is saturated.  The
weak energy condition then requires nonnegative energy density and
nonnegative transverse null combination.  The strong condition additionally
requires nonnegative transverse pressure, whereas the dominant condition
requires the magnitude of the transverse pressure not to exceed the energy
density.  For $1<\lambda<2$, the transverse NEC and SEC
outside a given radius require
\begin{align}
 r^2&\ge\frac{\lambda-1}{\lambda+1}a^2\cos^2\theta,\nonumber\\
 r^2&\ge(\lambda-1)a^2\cos^2\theta,
\end{align}
respectively.  To show how these analytical inequalities are realized outside
the horizon, Fig. \ref{fig:source} compares the exact axial energy density and the
ratio $p_\perp/\rho$ for the same representative KR parameters used in the
geometrical analysis below.

The left panel of Eq. \eqref{fig:source} shows that the effective density is largest
close to the event horizon and decays rapidly with radius. Increasing
$\lambda$ changes both its magnitude and falloff.  The right panel shows a
qualitative distinction that is less evident from the density alone:
$\lambda=0.5$ violates the transverse dominant-energy bound, $\lambda=1$
saturates it and $\lambda=1.5$ satisfies it over the displayed exterior
domain.  Thus a regular exterior density does not by itself guarantee that all
effective energy conditions hold.

\subsection{Relation to the complete Kalb-Ramond field equations}

The inverse construction proves that the geometry is supported by a symmetric,
conserved anisotropic source.  It does not, by itself, identify that source
with the stress tensor of a specific rotating KR field.  An exact solution of
the original nonminimally coupled theory additionally requires an explicit
$B_{\mu\nu}(r,\theta)$ satisfying simultaneously
\begin{align}
 G_{\mu\nu}&=\kappa T^{\KR}_{\mu\nu}[g,B],\nonumber\\
 \frac{\delta S}{\delta B_{\mu\nu}}&=0,\qquad
 B_{\mu\nu}B^{\mu\nu}=\mp b^2.
 \label{eq:fullKRcheck}
\end{align}
The Azreg-A\"{\i}nou algorithm fixes the rotating metric but does not supply
this two-form.  We therefore use the precise description ``rotating
power law geometry with an exact inverse Einstein source'' and regard it as a
phenomenological rotating extension of the static KR black hole.  All
geodesic and causal results below follow exactly from the metric. Thermodynamic
identifications that require the underlying Lagrangian are explicitly marked
as effective.

\section{Horizons and ergoregion}

The Killing horizons are positive roots of
\begin{equation}
 \Delta(r_H)=r_H^2+a^2-2Mr_H+\Upsilon r_H^q=0.
 \label{eq:horizon}
\end{equation}
When two roots exist, $r_-<r_+$ are the Cauchy and event horizons.  Extremality requires $\Delta=\Delta'=0$, giving
\begin{equation}
 \begin{aligned}
 M_e&=r_e+\frac q2\Upsilon r_e^{q-1},\\
 a_e^2&=r_e^2+(q-1)\Upsilon r_e^q.
 \end{aligned}
 \label{eq:extreme}
\end{equation}
The stationary limit surfaces (SLSs) obey $g_{tt}=0$:
\begin{equation}
 r_E^2-2Mr_E+a^2\cos^2\theta+\Upsilon r_E^q=0.
 \label{eq:sls}
\end{equation}
They meet the corresponding horizons at the poles.  The physical outer ergoregion is
\begin{equation}
 \begin{aligned}
 r_+&\le r\le r_E^+(\theta),\\
 \mathcal T_E(\theta)&=r_E^+(\theta)-r_+.
 \end{aligned}
\end{equation}

For a nonextremal static outer root $r_0$, define
\begin{equation}
 D_0=2r_0-2M+q\Upsilon r_0^{q-1}>0.
\end{equation}
At slow rotation,
\begin{equation}
 r_+=r_0-\frac{a^2}{D_0}+O(a^4),\qquad
 r_E^+=r_0-\frac{a^2\cos^2\theta}{D_0}+O(a^4),
\end{equation}
and, therefore,
\begin{equation}
 \mathcal T_E(\theta)=\frac{a^2\sin^2\theta}{D_0}+O(a^4).
 \label{eq:thickness}
\end{equation}
Thus, the thickness vanishes at the poles and is largest at the equator.  For
$\lambda=1$, the relevant surfaces and maximum thickness are elementary:
\begin{align}
 r_\pm&=M\pm\sqrt{M^2-a^2-\Upsilon},\nonumber\\
 r_E^\pm(\theta)&=M\pm\sqrt{M^2-\Upsilon-a^2\cos^2\theta},\label{eq:l1surfaces}\\
 \mathcal T_E^{\max}&=\sqrt{M^2-\Upsilon}
 -\sqrt{M^2-\Upsilon-a^2}.\nonumber
\end{align}

Two complementary measures quantify the size of the exterior ergoregion.  A
useful exact coordinate volume is
\begin{equation}
 V_{\rm coord}=\frac{2\pi}{3}\int_0^\pi\!\sin\theta
 \left\{[r_E^+(\theta)]^3-r_+^3\right\}\dd\theta,
 \label{eq:Vcoord}
\end{equation}
which has the slow rotation form
\begin{equation}
 V_{\rm coord}=\frac{8\pi r_0^2a^2}{3D_0}+O(a^4).
\end{equation}
The proper three-volume on a Boyer-Lindquist $t={\rm const}$ slice is instead
\begin{equation}
 V_{\rm proper}=2\pi\int_0^\pi\!\dd\theta
 \int_{r_+}^{r_E^+(\theta)}
 \sqrt{g_{rr}g_{\theta\theta}g_{\phi\phi}}\,\dd r,
 \label{eq:Vproper}
\end{equation}
and its integrable near-horizon behavior gives
\begin{equation}
 V_{\rm proper}=\frac{2\pi^2r_0^3}{D_0}|a|+O(a^2).
\end{equation}
Unlike the surface locations and thickness at fixed Boyer-Lindquist angle,
the proper three-volume depends on the chosen spatial slicing. Both volume
definitions are therefore stated explicitly.

\begin{table*}[t]
 \caption{Exact horizon and equatorial stationary-limit radii for $M=1$ and
 $\gamma=0.1$.  The last column is the maximum outer-ergoregion thickness
 $\mathcal T_E^{\max}=r_E^+(\pi/2)-r_+$.}
 \label{tab:ergoradii}
 \centering\small
 \begin{tabular}{ccccccr}
 \toprule
 $\lambda$ & $a$ & $r_-$ & $r_+$ & $r_E^-(\pi/2)$ & $r_E^+(\pi/2)$ & $\mathcal T_E^{\max}$\\
 \midrule
 0.5&0.1&0.39885&1.98211&0.39657&1.98726&0.00514\\
 0.5&0.5&0.46458&1.84897&0.39657&1.98726&0.13829\\
 0.5&0.9&0.80805&1.36973&0.39657&1.98726&0.61753\\
 1.0&0.1&0.05660&1.94340&0.05132&1.94868&0.00529\\
 1.0&0.5&0.19377&1.80623&0.05132&1.94868&0.14246\\
 1.0&0.9&0.70000&1.30000&0.05132&1.94868&0.64868\\
 1.5&0.1&0.00682&1.91424&0.00013&1.91954&0.00530\\
 1.5&0.5&0.15046&1.77673&0.00013&1.91954&0.14281\\
 1.5&0.9&0.66234&1.26971&0.00013&1.91954&0.64983\\
 \bottomrule
 \end{tabular}
\end{table*}

\begin{figure*}[t]
 \centering
 \includegraphics[width=0.96\textwidth]{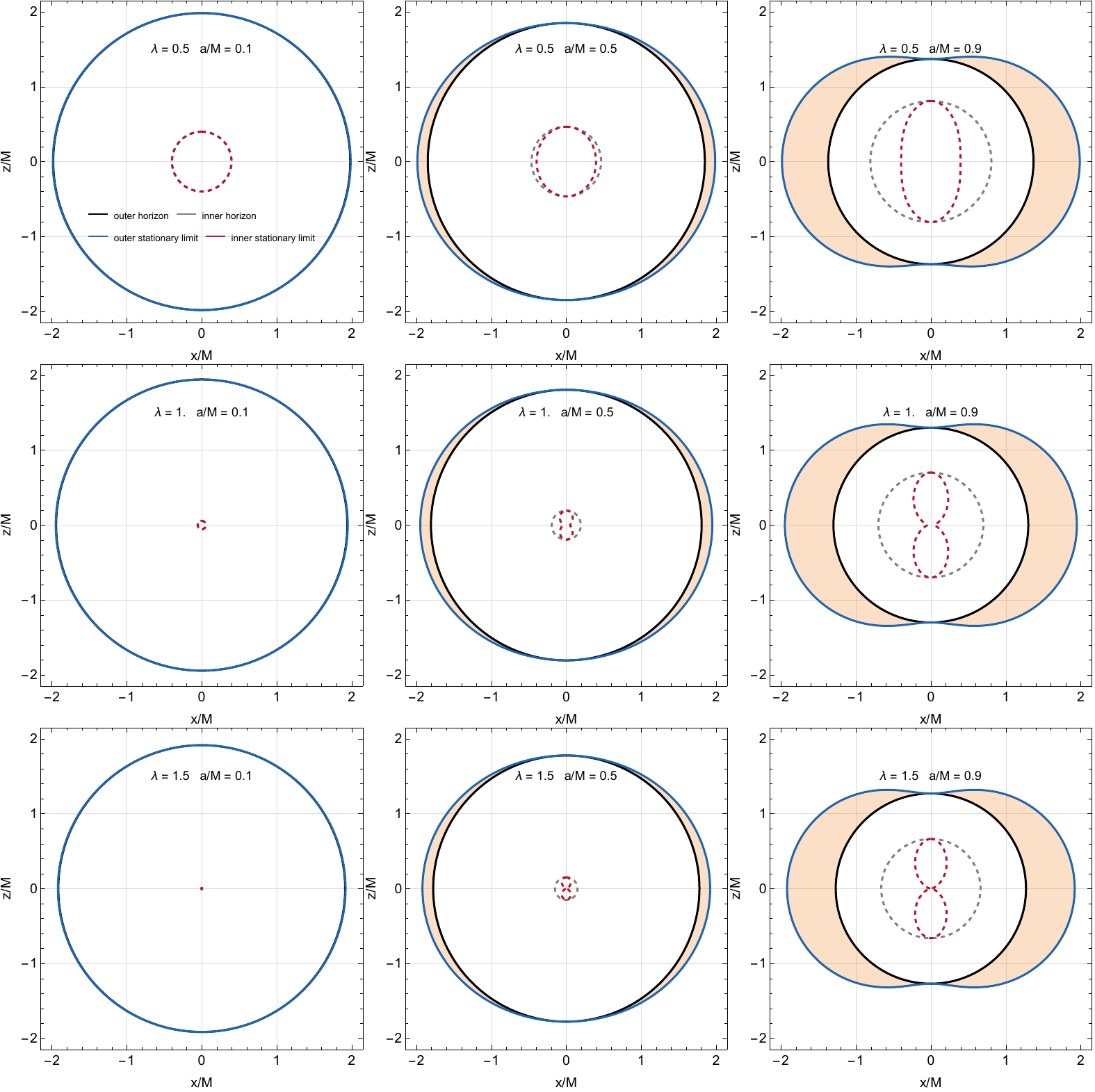}
 \caption{\emph{Exact meridional sections of the inner and outer horizons and stationary-limit surfaces.  The shaded band is the outer ergoregion.  Increasing spin produces the dominant widening, while varying $\lambda$ strongly reorganizes the inner surfaces.}}
 \label{fig:horizons}
\end{figure*}

The exact numerical surfaces in Fig. \ref{fig:horizons} use $M=1$ and
$\gamma=0.1$ and, so, no small-deformation or slow-spin approximation is used.  The
figure is introduced here to visualize simultaneously the inner and outer
roots listed in Table \ref{tab:ergoradii}, their angular stationary-limit
counterparts and the finite region between the event horizon and outer SLS.

From Fig.  \ref{fig:horizons}, it appears clear that the horizon remains circular in a
meridional section because $\Delta=0$ is independent of $\theta$, whereas each
SLS is oblate and joins its corresponding horizon at the poles.  Spin is the
principal control on the width of the outer ergoregion.  The KR parameter has
a comparatively modest effect on the outer surface but can substantially
rearrange the inner horizon and inner SLS. The latter are therefore more
sensitive checks of the power law deformation, although they lie behind the
event horizon.

\begin{figure*}[t]
 \centering
 \includegraphics[width=1.\textwidth]{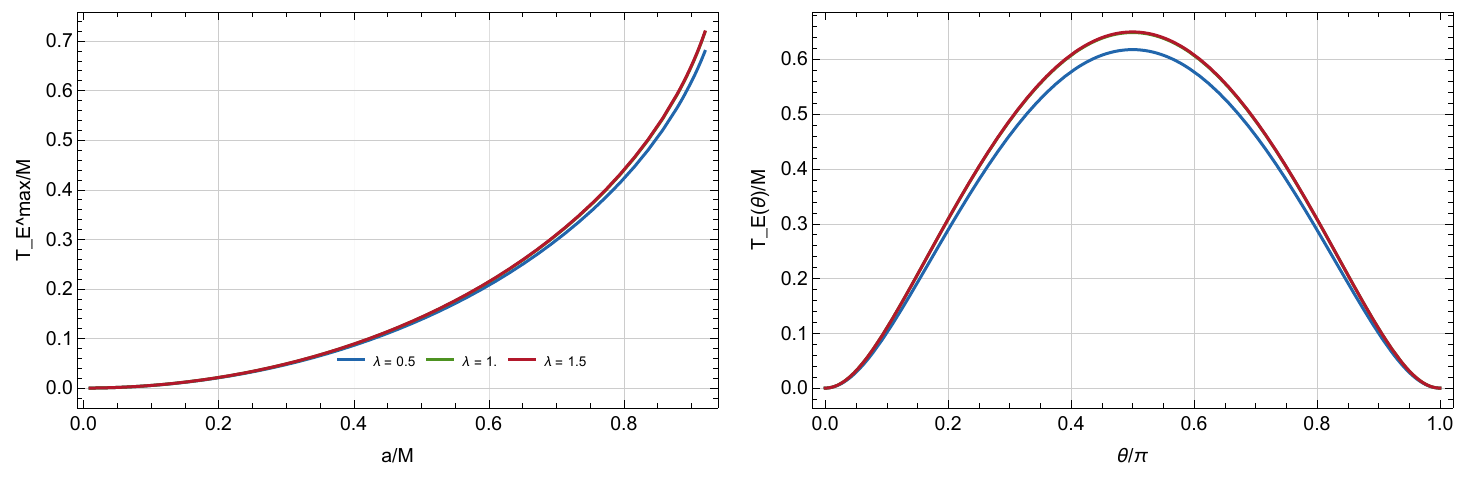}
 \caption{\emph{Exact outer-ergoregion thickness for $M=1$ and $\gamma=0.1$.
 The left panel shows the maximum equatorial thickness as a function of spin.
 The right panel shows the full angular profile at $a/M=0.9$, where the thickness
 vanishes at the poles, is symmetric about the equatorial plane and reaches
 its maximum at $\theta=\pi/2$.  The coordinate and proper volumes remain
 quantified by Eqs. \eqref{eq:Vcoord} and \eqref{eq:Vproper}.}}
 \label{fig:ergoquant}
\end{figure*}

While Fig. \ref{fig:horizons} displays the surfaces themselves, Fig.
\ref{fig:ergoquant} isolates two observable geometric measures of the outer
ergoregion namely its maximum equatorial thickness as spin varies and its complete
angular thickness at fixed rapid rotation.

The nearly coincident low-spin curves in Fig. \ref{fig:ergoquant} confirm the
$\mathcal T_E\propto a^2$ behavior of Eq. \eqref{eq:thickness}.  Their separation
becomes visible only toward rapid rotation, where the $\lambda=1.5$ branch is
slightly thicker than the $\lambda=0.5$ branch.  The angular panel also shows
that this enhancement is concentrated around the equator and disappears at
both poles.  Consequently, the integrated ergoregion volumes in Eqs.
\eqref{eq:Vcoord} and \eqref{eq:Vproper} respond more strongly at high spin than in the
slow rotation regime.

\section{Null geodesics and photon region}

For a photon, the separable Hamilton-Jacobi action is
\begin{equation}
 \mathcal S=-Et+L_z\phi+S_r(r)+S_\theta(\theta).
\end{equation}
Writing $\xi=L_z/E$ and $\eta=\mathcal Q/E^2$, the potentials are
\begin{align}
 \frac{\mathcal R}{E^2}&=(r^2+a^2-a\xi)^2-\Delta[(\xi-a)^2+\eta],
 \label{eq:Rpot}\\
 \frac{\Theta}{E^2}&=\eta+a^2\cos^2\theta-\xi^2\cot^2\theta.
\end{align}
The first-order equations include
\begin{align}
 \Sigma\dot r&=\pm\sqrt{\mathcal R},\nonumber\\
 \Sigma\dot\phi&=\frac{a[E(r^2+a^2)-aL_z]}{\Delta}
 +\frac{L_z}{\sin^2\theta}-aE.
 \label{eq:firstorder}
\end{align}
Spherical photon orbits satisfy $\mathcal R=\mathcal R'=0$ \cite{Teo2003}, giving
\begin{align}
 \xi_c(r_p)&=\frac{(r_p^2+a^2)\Delta'_p-4r_p\Delta_p}{a\Delta'_p},
 \label{eq:xi}\\
 \eta_c(r_p)&=\frac{16r_p^2\Delta_p}{(\Delta'_p)^2}-[\xi_c(r_p)-a]^2.
 \label{eq:eta}
\end{align}
The relevant capture boundary additionally obeys $r_p>r_+$, $\Theta\ge0$ and the radial instability condition $\mathcal R''(r_p)>0$ in the convention of Eq. \eqref{eq:Rpot}.  Equatorial prograde and retrograde radii are the solutions of
\begin{equation}
 r_p\Delta'_p-4\Delta_p=\pm4a\sqrt{\Delta_p}.
\end{equation}
At $a=0$, this reduces to
\begin{equation}
 r_c-3M+\left(1+\frac1\lambda\right)\Upsilon r_c^{1-2/\lambda}=0.
 \label{eq:staticph}
\end{equation}
For slow rotation define $\Delta_0=r^2-2Mr+\Upsilon r^q$ and
$A(r)=r\Delta_0'(r)-4\Delta_0(r)$.  If $A(r_c)=0$ is a simple static photon
root, the two equatorial branches split as
\begin{equation}
 r_{\rm ph}^{\rm pro/retro}=r_c\pm
 \frac{4a\sqrt{\Delta_0(r_c)}}{A'(r_c)}+O(a^2),
 \label{eq:slowphoton}
\end{equation}
where the upper sign denotes the prograde branch for $a>0$.  As an explicit
check, Kerr has $r_c=3M$, $A'(r_c)=-6M$ and
$\Delta_0(r_c)=3M^2$, so this equation gives
$r_{\rm ph}^{\rm pro/retro}=3M\mp2a/\sqrt3+O(a^2)$, with the physically
smaller prograde radius.
For $\lambda=1$ the exact implicit equation simplifies to
\begin{equation}
 r^2-3Mr+2\Upsilon\pm2a\sqrt{Mr-\Upsilon}=0.
 \label{eq:l1photon}
\end{equation}

\subsection{Photon ergosurface gap}

The equatorial separation between an unstable photon orbit and the outer
stationary-limit surface provides a direct measure of how the optical capture
region is positioned relative to the ergoregion.  We define the two signed
gaps by
\begin{align}
 &\mathcal G_{\rm pro}=r_{\rm ph}^{\rm pro}-r_E^+(\pi/2),\nonumber\\
 \,\\
 &\mathcal G_{\rm retro}=r_{\rm ph}^{\rm retro}-r_E^+(\pi/2).\nonumber
 \label{eq:photongap}
\end{align}
A positive value means that the orbit lies outside the ergoregion, whereas a
negative value means that it lies between the event horizon and the outer
stationary-limit surface.  Notice that the equatorial SLS is independent of
spin and coincides with the outer root $r_0$ of the static seed equation,
while both photon radii retain a strong spin dependence.

Combining Eq. \eqref{eq:slowphoton} with $r_E^+(\pi/2)=r_0$ gives the useful
slow rotation approximation
\begin{equation}
 \begin{aligned}
 \mathcal G_{\rm pro}
 &\simeq r_c-r_0
 +\frac{4a\sqrt{\Delta_0(r_c)}}{A'(r_c)}+O(a^2),\\[2pt]
 \mathcal G_{\rm retro}
 &\simeq r_c-r_0
 -\frac{4a\sqrt{\Delta_0(r_c)}}{A'(r_c)}+O(a^2).
 \end{aligned}
 \label{eq:gapslow}
\end{equation}
where $r_c$ is the static photon radius and $r_0$ is the static outer horizon.
This expression separates the static KB displacement $r_c-r_0$
from the leading frame-dragging splitting. More precisely, the numerical results below use
the exact photon-orbit relation.

\begin{table*}[t]
 \caption{Exact equatorial photon ergosurface gaps for $M=1$ and
 $\gamma=0.1$, with Kerr included as the $\gamma=0$ reference.  The prograde
 orbit enters the ergoregion when $\mathcal G_{\rm pro}<0$, whereas the retrograde
 orbit remains outside it for all displayed cases.}
 \label{tab:photongap}
 \centering\small
 \begin{tabular}{llrrrrr}
 \toprule
 $a/M$ & Model & $r_E^+(\pi/2)$ & $r_{\rm ph}^{\rm pro}$ &
 $\mathcal G_{\rm pro}$ & $r_{\rm ph}^{\rm retro}$ &
 $\mathcal G_{\rm retro}$\\
 \midrule
 0.1 & Kerr          &2.00000&2.88219& 0.88219&3.11335&1.11335\\
     & $\lambda=0.5$&1.98726&2.86975& 0.88249&3.10311&1.11586\\
     & $\lambda=1$  &1.94868&2.81174& 0.86305&3.04704&1.09836\\
     & $\lambda=1.5$&1.91954&2.76383& 0.84429&2.99723&1.07769\\
 \addlinespace
 0.5 & Kerr          &2.00000&2.34730& 0.34730&3.53209&1.53209\\
     & $\lambda=0.5$&1.98726&2.32565& 0.33839&3.52459&1.53733\\
     & $\lambda=1$  &1.94868&2.26145& 0.31277&3.47132&1.52264\\
     & $\lambda=1.5$&1.91954&2.21816& 0.29862&3.41814&1.49860\\
 \addlinespace
 0.9 & Kerr          &2.00000&1.55785&$-0.44215$&3.91027&1.91027\\
     & $\lambda=0.5$&1.98726&1.46624&$-0.52101$&3.90438&1.91712\\
     & $\lambda=1$  &1.94868&1.37843&$-0.57025$&3.85311&1.90443\\
     & $\lambda=1.5$&1.91954&1.34389&$-0.57564$&3.79696&1.87743\\
 \bottomrule
 \end{tabular}
\end{table*}

\begin{figure*}[t]
 \centering
 \includegraphics[width=1.\textwidth]{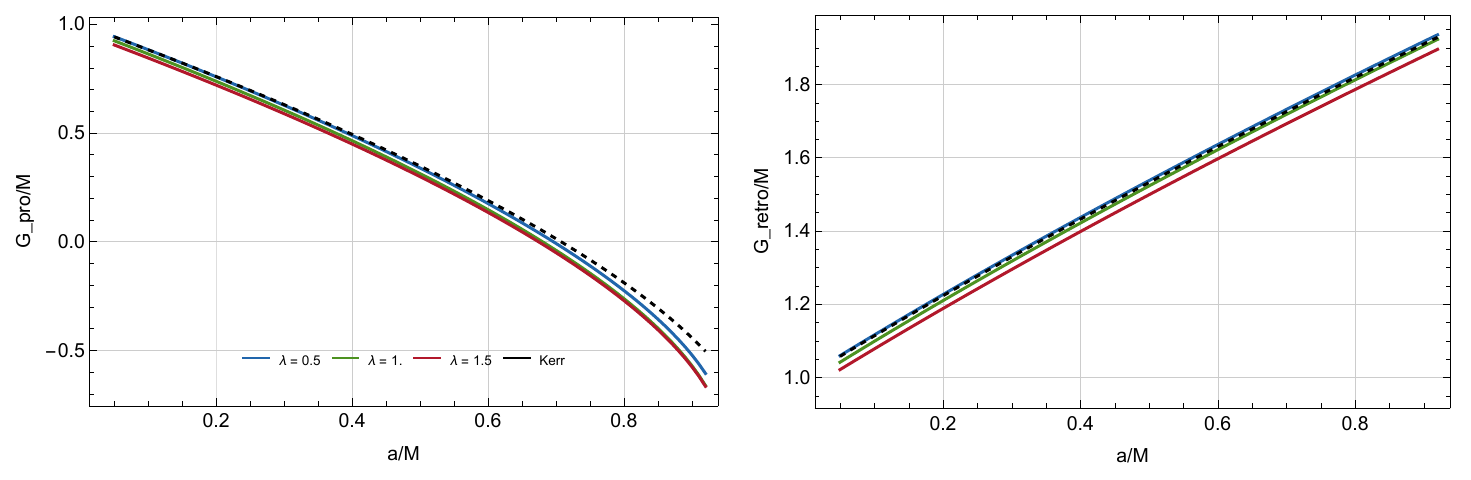}
 \caption{\emph{Exact signed photon ergosurface gaps as functions of spin.  The
 zero crossing in the left panel marks entry of the prograde photon orbit into
 the ergoregion.  The KB deformation shifts this transition to lower
 spin, whereas the retrograde orbit remains outside the ergoregion.}}
 \label{fig:photongap}
\end{figure*}

The deformation moves both the equatorial SLS and the photon orbits inward,
but not by equal amounts.  At $a/M=0.5$, increasing $\lambda$ from $0.5$ to
$1.5$ decreases the prograde gap from $0.33839M$ to $0.29862M$.  At
$a/M=0.9$ the prograde orbit is already inside the ergoregion and the
deformation increases the penetration depth from $0.44215M$ in Kerr to
$0.52101M$, $0.57025M$ and $0.57564M$ for $\lambda=0.5$, $1$ and $1.5$,
respectively.  The exact crossing $\mathcal G_{\rm pro}=0$ occurs at
the following critical spins (the KB entries use $\gamma=0.1$):
\begin{equation}
 \left(\frac{a}{M}\right)_{\mathcal G_{\rm pro}=0}\simeq
 \begin{cases}
 0.70711, & \text{Kerr},\\
 0.69581, & \lambda=0.5,\\
 0.67983, & \lambda=1,\\
 0.67390, & \lambda=1.5.
 \end{cases}
 \label{eq:gapcriticalspin}
\end{equation}
This turns out to be an important qualitative result, namely positive $\Upsilon$ causes the
prograde photon orbit to enter the ergoregion at a lower spin. Conversely,
the retrograde photon orbit remains outside the ergoregion throughout the
investigated black-hole domain and its gap is comparatively insensitive, as
shown in Fig. \ref{fig:photongap}.

The left panel of Fig. \ref{fig:photongap} converts the critical-spin values in Eq.
\eqref{eq:gapcriticalspin} into a continuous geometric picture, where each zero
crossing marks the transition from a prograde orbit outside the SLS to one
inside the ergoregion.  Increasing $\lambda$ shifts this crossing leftward and
deepens the negative high-spin gap.  Conversely, the positive and slowly
varying retrograde curves in the right panel show that the counter-rotating
photon orbit remains well separated from the ergoregion.  The KR deformation
therefore acts asymmetrically on the two photon branches once frame dragging
is present.

\section{Shadow and critical curve deviation}

\begin{figure*}[t]
 \centering
 \includegraphics[width=1.\textwidth]{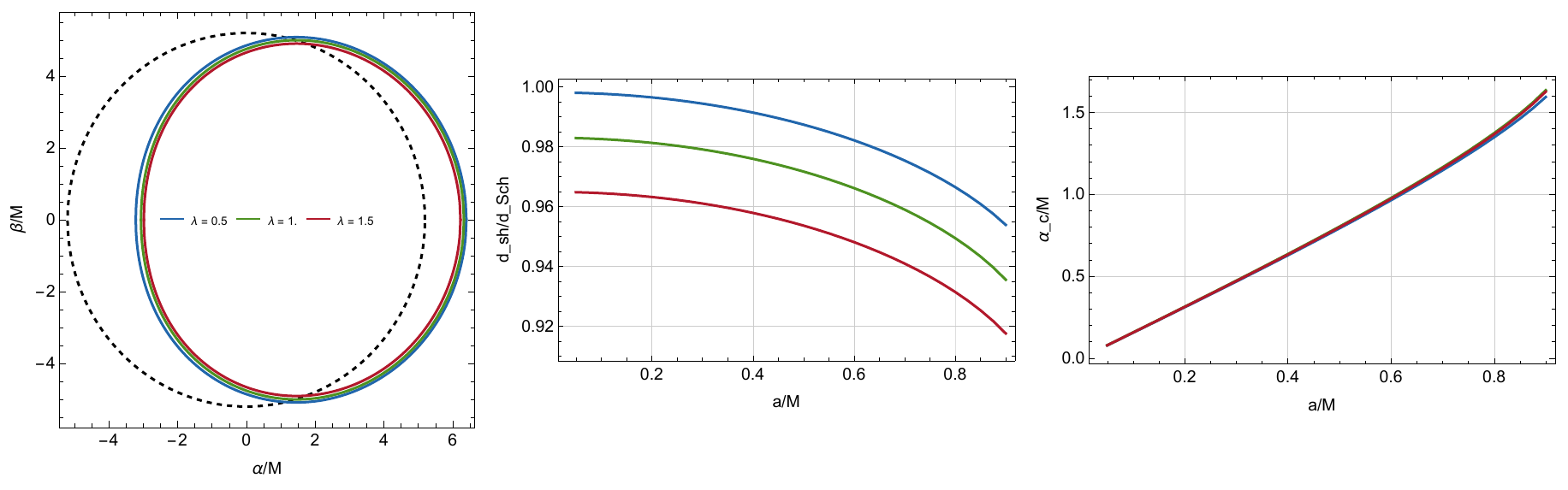}
 \caption{\emph{Complementary exact critical curve checks at $\gamma=0.1$ and
 $\theta_o=50^\circ$.  Left: the rapidly rotating shape compared with the
 Schwarzschild circle.  Center: the area equivalent diameter normalized by
 $d_{\rm Sch}=6\sqrt3M$.  Right: the horizontal displacement, which vanishes
 for Schwarzschild.  The panels show shape, size and position.}}
 \label{fig:shadow}
\end{figure*}

For a distant observer at inclination $\theta_o$, the critical curve is
\cite{PerlickTsupko2022}
\begin{align}
 \alpha(r_p)&=-\xi_c(r_p)\csc\theta_o,\nonumber\\
 \beta(r_p)&=\pm\sqrt{\eta_c(r_p)+a^2\cos^2\theta_o
 -\xi_c^2(r_p)\cot^2\theta_o}.
 \label{eq:celestial}
\end{align}
Its area and area equivalent diameter are
\begin{equation}
\begin{aligned}
 &A_{\rm sh}=2\int\beta\frac{\dd\alpha}{\dd r_p}\dd r_p,\\
 &d_{\rm sh}=2\sqrt{\frac{|A_{\rm sh}|}{\pi}}.
 \label{eq:area}
\end{aligned}
\end{equation}
The area equivalent diameter and horizontal displacement are
compared in Fig. \ref{fig:shadow}.  Positive $\Upsilon$ primarily decreases the enclosed
area, whereas rotation produces the leading horizontal displacement.  The
Schwarzschild values provide a common normalization, i.e., a centered circle with
$d_{\rm Sch}=6\sqrt3M$.

The three plots, drawn in Fig. \ref{fig:shadow}, separate effects that would be
partly degenerate in a single  comparison.  At $a/M=0.9$, the left
panel shows that the KR curves retain the familiar displaced, asymmetric
rotating outline.  The middle panel demonstrates a monotonic reduction of the
area equivalent diameter with both spin and increasing $\lambda$ for the
chosen positive deformation, while the right panel shows that the centroid
displacement is dominated by rotation and depends only weakly on $\lambda$.
Hence size is the cleaner diagnostic of the KR correction in this parameter
range, whereas horizontal displacement primarily constrains spin.

The fractional deviation from the Schwarzschild critical diameter is
\begin{equation}
 \delta=\frac{d_{\rm sh}}{6\sqrt3M}-1.
 \label{eq:delta}
\end{equation}
For a static black hole and small $\gamma$,
\begin{align}
 \frac{r_c}{M}&=3-\left(1+\frac1\lambda\right)
 3^{1-2/\lambda}\gamma+O(\gamma^2),\\
 \frac{d_{\rm sh}}{M}&=6\sqrt3\left[1-\frac32\,3^{-2/\lambda}\gamma
 \right]+O(\gamma^2).
\end{align}
Including the leading areal Kerr correction gives the useful approximation
\begin{equation}
 \delta\simeq-\frac{1+\cos^2\theta_o}{36}a_*^2
 -\frac32\,3^{-2/\lambda}\gamma,
 \label{eq:deltaapprox}
\end{equation}
with neglected terms $O(a_*^4,\gamma^2,a_*^2\gamma)$.

The EHT compatible intervals used here are $\delta=-0.01\pm0.17$ for M87$^*$ and $\delta=-0.08\pm0.09$ (VLTI) or $-0.04^{+0.09}_{-0.10}$ (Keck) for Sgr~A$^*$ \cite{EHTM87VI,EHTSgrVI,Walia2022}.  We restrict this observational interpretation to $0<\lambda<2$, for which the KR correction decays faster than $1/r$ and $M$ has the standard asymptotic mass normalization.  Conditional on $\gamma=0.1$, $\theta_o=17^\circ$ for M87$^*$ and $50^\circ$ for Sgr~A$^*$, Table \ref{tab:EHT} reports the value of $\lambda$ obtained by solving the \emph{exact} area equivalent critical curve defined above,
$\delta_{\rm exact}(\lambda;\gamma,a_*,\theta_o)=\delta_{\rm obs}^{\rm central}$.
Thus each entry is an illustrative central-value inversion at fixed spin, not
a predicted shadow diameter, confidence interval, or independent measurement
of $\lambda$.  EHT actually constrains the joint space
$(\lambda,\gamma,a_*,\theta_o)$.

\begin{table}[t]
 \caption{Values of $\lambda$ that reproduce the central observed fractional
 critical curve deviation from the exact area equivalent critical curve,
 after fixing $\gamma=0.1$, $a_*$ and the stated inclination.  The adopted
 domain is $0<\lambda<2$.  These are illustrative one-parameter inversions,
 not best-fit estimates.  A dash means that no central-value solution exists
 inside this domain, although the model can still lie within the observational
 interval.}
 \label{tab:EHT}
 \centering
 \begin{tabular}{lccc}
 \toprule
 Case & $a_*=0.1$ & $a_*=0.5$ & $a_*=0.9$\\
 \midrule
 M87$^*$ & $0.79$ & - & -\\
 Sgr~A$^*$ VLTI & - & - & $1.42$\\
 Sgr~A$^*$ Keck & $1.64$ & $1.31$ & Kerr limit\\
 \bottomrule
 \end{tabular}
\end{table}

The formal exact central-value roots $\lambda=3.67$ and $2.93$ for the first
two VLTI spin choices have therefore been omitted, as they lie outside the
adopted asymptotic domain.  The value $1.42$ at $a_*=0.9$ is an exact
critical curve result. Conversely, inserting the same parameters into the
leading formula, Eq.  \eqref{eq:deltaapprox},  gives approximately $1.94$.  This
comparison quantifies why the approximation must not be used for high-spin
parameter inversion.  One-standard-deviation compatibility should likewise
be evaluated only after intersecting the observational region with
$0<\lambda<2$.  Within that interval the broad EHT bands leave a strong
$(\gamma,\lambda,a_*,\theta_o)$ degeneracy, so the table must not be read as
a stand-alone measurement or exclusion of $\lambda$.

\begin{figure*}[t]
 \centering
 \includegraphics[width=1.\textwidth]{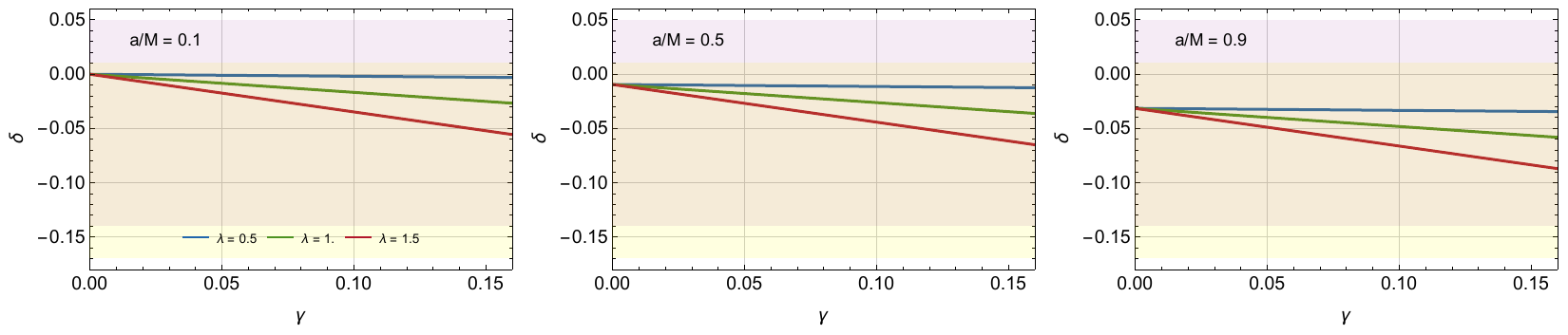}
 \caption{\emph{Leading fractional critical curve deviation from Eq. \eqref{eq:deltaapprox}.  The shaded bands show the Sgr~A$^*$ one-standard-deviation intervals.  Exact high-spin inference must use the full critical curve area.}}
 \label{fig:delta}
\end{figure*}

The dependence behind these bounds is displayed explicitly in Fig.
\ref{fig:delta}, where the approximation in Eq. \eqref{eq:deltaapprox} is evaluated as
a function of $\gamma$ at each fixed spin.  Plotting the observational bands,
together with the theoretical curves, makes the spin deformation degeneracy
and the different slopes of the three $\lambda$ regions quite clear.

For positive $\gamma$, every curve in Fig. \ref{fig:delta} moves toward a smaller
critical curve diameter and the magnitude of the slope increases with
$\lambda$.  Increasing spin adds a negative offset, so the high-spin branches
approach the lower edge of the observational interval sooner.  The figure also
explains why the central value inversions in Table \ref{tab:EHT} should not be
treated as unique parameter measurements, i.e.,  comparable deviations can be
generated by changing either $a_*$, $\gamma$, or $\lambda$.

\section{Surface gravity and geometric response functions}

The horizon equation gives
\begin{equation}
 M=\frac{r_+^2+a^2+\Upsilon r_+^q}{2r_+}.
\end{equation}
Assuming the area entropy, the geometric quantities are
\begin{align}
 S_A&=\pi(r_+^2+a^2),\qquad
 \Omega_H=\frac{a}{r_+^2+a^2},\\
 T_H&=\frac{\Delta'(r_+)}{4\pi(r_+^2+a^2)}
 =\frac{r_+^2-a^2+(q-1)\Upsilon r_+^q}
 {4\pi r_+(r_+^2+a^2)}.
 \label{eq:TH}
\end{align}
For $J=aM$, direct differentiation yields
\begin{equation}
\begin{split}
 \dd M-T_H\dd S_A-\Omega_H\dd J={}&
 \frac{r_+^{q+1}}{2(r_+^2+a^2)}\dd\Upsilon\\
 &-\frac{aq\Upsilon r_+^{q-1}}{2(r_+^2+a^2)}\dd a.
 \label{eq:firstlawresidual}
\end{split}
\end{equation}
On the Kerr family the residual vanishes when $\Upsilon=0$ and
$\dd\Upsilon=0$.  At $q=0$ ($\lambda=1$), however, it reduces to
$r_+\dd\Upsilon/[2(r_+^2+a^2)]$ and vanishes only in the fixed-$\Upsilon$
ensemble. If $\Upsilon$ varies, a charge-like work term is required.  Thus a
general-$\lambda$ interpretation based simultaneously on $S_A$, $J=aM$ and
$T_H$ remains geometric until the Wald entropy and physical charges of the
rotating KR theory are established \cite{Wald1993}.

At fixed $J$, $\Upsilon$ and $\lambda$,
\begin{align}
 \left(\frac{\dd a}{\dd r_+}\right)_J
 &=-\frac{a r_+M_r}{r_+M+a^2},\nonumber\\
 M_r&=\frac12-\frac{a^2}{2r_+^2}
 +\frac{q-1}{2}\Upsilon r_+^{q-2}.
\end{align}
We now consider  $A=r_+^2+a^2$ and
\begin{align}
 N&=r_+-\frac{a^2}{r_+}+(q-1)\Upsilon r_+^{q-1},\nonumber\\
 T_H&=\frac{N}{4\pi A},\qquad N_a=-\frac{2a}{r_+},\nonumber\\
 N_r&=1+\frac{a^2}{r_+^2}+(q-1)^2\Upsilon r_+^{q-2},\nonumber\\
 (T_H)_r&=\frac{A N_r-2r_+N}{4\pi A^2},\nonumber\\
 (T_H)_a&=\frac{-2aA/r_+-2aN}{4\pi A^2}.
 \label{eq:Tpartials}
\end{align}
Consequently,
\begin{align}
 \left(\frac{\dd T_H}{\dd r_+}\right)_J
 &=(T_H)_r+(T_H)_a\left(\frac{\dd a}{\dd r_+}\right)_J,\nonumber\\
 \left(\frac{\dd S_A}{\dd r_+}\right)_J
 &=2\pi\left[r_++a\left(\frac{\dd a}{\dd r_+}\right)_J\right],
 \label{eq:fixedJderivatives}
\end{align}
which supply a fully explicit one dimensional response calculation at each fixed
$J$. On the other hand, if one enforces a formal first law with $S_A$ and $J$ as the
thermodynamic variables, the conjugate temperature, $T_{\rm th}=\left(\frac{\partial M}{\partial S_A}\right)_{J,\Upsilon}$, becomes
\begin{equation}
 T_{\rm th}=
 \frac{M M_r}{2\pi(r_+M+a^2-a^2M_r)}.
\end{equation}
For general $\lambda$, $T_{\rm th}\ne T_H$. This mismatch is precisely the
consistency warning encoded by Eq. \eqref{eq:firstlawresidual}.
We therefore define the fixed-$J$ geometric thermal response and area-based
geometric potential
\begin{align}
 \mathcal R_J^{(T)}&=T_H\frac{(\dd S_A/\dd r_+)_J}{(\dd T_H/\dd r_+)_J},
 \label{eq:CJ}\\
 \mathcal G_A&=M-T_HS_A
 =\frac{r_+^2+3a^2+(3-q)\Upsilon r_+^q}{4r_+}.
 \label{eq:F}
\end{align}
The sign of $\mathcal R_J^{(T)}$ records the orientation of the
area-surface-gravity-temperature response and its divergence marks a
turning point where $(\dd T_H/\dd r_+)_J=0$.  Neither this sign nor
$\mathcal G_A$ by itself is a thermodynamic stability criterion. Indeed, to this end, one would require the Wald entropy, the correct conserved charges and their conjugate
potentials.

\section{Rotational energy extraction}

The horizon-crossing condition for a particle is
\begin{equation}
 E-\Omega_HL\ge0.
\end{equation}
Negative-energy states can exist inside the ergoregion.  If a fragment with $E_{\rm in}<0$ enters the black hole, the escaping fragment gains $-E_{\rm in}$.  The ideal equatorial splitting efficiency is
\begin{equation}
 \eta_P^{\max}=\frac12\left(\sqrt{1+\frac{a^2}{r_+^2}}-1\right).
\end{equation}
For neutral waves the superradiant interval is
\begin{equation}
 0<\omega<m\Omega_H.
\end{equation}
An effective reversible spin-down at fixed $S_A$, $\Upsilon$ and $\lambda$ preserves $R_A=\sqrt{r_+^2+a^2}$.  Its maximum geometric energy estimate is
\begin{equation}
 E_{\rm rot}^{\max}=M_i-\left[\frac{R_A}{2}+\frac{\Upsilon}{2}R_A^{q-1}\right].
\end{equation}
Because of Eq. \eqref{eq:firstlawresidual}, this should not be promoted to an exact KR thermodynamic theorem for any $\lambda$.

\section{Rotating weak deflection}

On the equatorial plane, dividing the azimuthal equation in \eqref{eq:firstorder} by the radial equation gives
\begin{equation}
 \frac{\dd\phi}{\dd r}=\pm
 \frac{a(r^2+a^2-a\xi)/\Delta+\xi-a}
 {\sqrt{(r^2+a^2-a\xi)^2-\Delta(\xi-a)^2}}.
 \label{eq:dphidr}
\end{equation}
If $r_0$ is the turning point, the exact asymptotic equatorial bending is
\begin{equation}
 \hat\alpha=2\int_{r_0}^{\infty}\left|\frac{\dd\phi}{\dd r}\right|\dd r-\pi.
 \label{eq:exactbend}
\end{equation}
For weak fields, the leading static power law coefficient is
\begin{equation}
 C_\lambda=\left(1+\frac1\lambda\right)\sqrt\pi
 \frac{\Gamma(3/2+1/\lambda)}{\Gamma(2+1/\lambda)},
\end{equation}
and the expansion begins as
\begin{equation}
 \hat\alpha_\pm=\frac{4M}{b}\pm\frac{4Ma}{b^2}
 +\frac{15\pi M^2}{4b^2}-C_\lambda\frac{\Upsilon}{b^{2/\lambda}}
 +O(a\Upsilon,M\Upsilon,a^2).
 \label{eq:weakgeneric}
\end{equation}
Equation \eqref{eq:weakgeneric} displays separate leading effects, not the complete rotating KR series.  For $\lambda=1$, the mixed contribution starts explicitly as
\begin{equation}
\begin{split}
 \hat\alpha_{\rm pro}={}&\frac{4M}{b}-\frac{4Ma}{b^2}
 +\frac{15\pi M^2-3\pi\Upsilon}{4b^2}\\
 &+\frac{128M^3}{3b^3}-\frac{10\pi M^2a}{b^3}
 +\frac{4Ma^2}{b^3}\\
 &+\frac{\pi a\Upsilon}{2b^3}
 -\frac{32M\Upsilon}{3b^3}+\cdots,
\end{split}
\end{equation}
with odd powers of $a$ reversed for retrograde propagation \cite{Kumar2020}.
The leading KB deformation contributions, linear in $\Upsilon$, for
the three representative values of $\lambda$ are
\begin{equation}
 \hat\alpha_{\rm KR}^{(1)}=\begin{cases}
 -15\pi\Upsilon/(16b^4),&\lambda=0.5,\\
 -3\pi\Upsilon/(4b^2),&\lambda=1,\\
 -\dfrac53\sqrt\pi\dfrac{\Gamma(13/6)}{\Gamma(8/3)}
 \Upsilon/b^{4/3},&\lambda=1.5.
 \end{cases}
 \label{eq:threebend}
\end{equation}
Mixed spin-KB contributions enter at order
$a\Upsilon/b^{1+2/\lambda}$ and require the full rotating expansion when
quantitative high-accuracy lensing is intended.

\section{Weak lensing magnification}

The scalar equatorial equation is sufficient for locating equatorial images,
but not for determining their physical flux magnification once spin breaks
circular symmetry.  We therefore use the two dimensional thin lens map and
its full Jacobian, following the weak deflection Kerr construction of
Refs.~\cite{SerenoDeLuca2006,WernerPetters2007}.  Let
$\boldsymbol{x}=\boldsymbol{\theta}/\theta_E$ and
$\boldsymbol{y}=\boldsymbol{\beta}_s/\theta_E$, where
\begin{equation}
\begin{aligned}
 &\theta_E=\sqrt{\frac{4M D_{LS}}{D_LD_S}},\\
 &\epsilon=\frac{M}{D_L\theta_E},\\
 &p=\frac{2}{\lambda}.
 \end{aligned}
\end{equation}
The projected spin displacement and the KR amplitude are
\begin{equation}
 \boldsymbol d=\epsilon\frac{\boldsymbol a\times\hat{\boldsymbol k}}{M},
 \qquad c=\frac{15\pi}{16}\epsilon,\qquad
 h=\frac{C_\lambda}{4}\gamma\epsilon^{p-1},
 \label{eq:lensparameters}
\end{equation}
where $\hat{\boldsymbol k}$ is the unperturbed propagation direction and
$|\boldsymbol d|=a_*\epsilon\sin\theta_o$.  Retaining the standard
post-Newtonian mass correction together with the leading spin and KR terms,
the dimensionless lens map is
\begin{equation}
\begin{split}
 \boldsymbol y={}&\boldsymbol x-\frac{\boldsymbol x}{r^2}
 -c\frac{\boldsymbol x}{r^3}
 +\frac{\boldsymbol d}{r^2}-2\frac{(\boldsymbol d\!\cdot\!\boldsymbol x)\boldsymbol x}{r^4}
 +h\frac{\boldsymbol x}{r^{p+1}},
 \label{eq:lensmap2d}
\end{split}
\end{equation}
with $r=|\boldsymbol x|$.

The two spin terms are the leading expansion of a Schwarzschild lens displaced
in the projected equatorial direction.  When source, image and
$\boldsymbol d$ are collinear, \eqref{eq:lensmap2d} reduces to the previous
one dimensional image-position equation, but the transverse derivative is
retained below.

The Jacobian $\mathcal A_{ij}=\partial y_i/\partial x_j$ is
\begin{equation}
\begin{split}
 \mathcal A_{ij}={}&\delta_{ij}-\frac{\delta_{ij}}{r^2}
 +\frac{2x_i x_j}{r^4}
 -c\left(\frac{\delta_{ij}}{r^3}-\frac{3x_i x_j}{r^5}\right)\\
 &-\frac{2(d_i x_j+x_i d_j)}{r^4}\\
 &-\frac{2u\delta_{ij}}{r^4}+\frac{8u x_i x_j}{r^6}
 +h\frac{\delta_{ij}}{r^{p+1}}\\
 &-h(p+1)\frac{x_i x_j}{r^{p+3}},
 \label{eq:lensjacobian}
\end{split}
\end{equation}
with $u=\boldsymbol d\!\cdot\!\boldsymbol x$.

The physical signed and unresolved magnifications are therefore
\begin{equation}
 \mu_i=\frac{1}{\det\mathcal A(\boldsymbol x_i)},
 \label{eq:mu2d}
\end{equation}
with $ \mu_{\rm tot}=\sum_i|\mu_i|$.

A negative $\mu_i$ denotes reversed parity, not negative flux.  For example,
$\mu_{\rm tot}=2.1$ means that the unresolved lensed source flux is $2.1$
times its unlensed value.

For comparison, the Schwarzschild image positions and signed magnifications
for a source on the first screen axis are
\begin{equation}
 x_{0\pm}=\frac12\left(y\pm\sqrt{y^2+4}\right),\qquad
 \mu_{0\pm}=\frac12\pm\frac{y^2+2}{2y\sqrt{y^2+4}}.
 \label{eq:schwl}
\end{equation}
For a general source, the angular image separation is
$\Delta\boldsymbol\theta=\theta_E(\boldsymbol x_+-\boldsymbol x_-)$ and the
unresolved astrometric centroid and centroid shift are
\begin{equation}
 \begin{aligned}
 &\boldsymbol\theta_{\rm c}=\theta_E
 \frac{\sum_i|\mu_i|\boldsymbol x_i}{\sum_i|\mu_i|},\\
 &\delta\boldsymbol\theta_{\rm c}=\boldsymbol\theta_{\rm c}
 -\theta_E\boldsymbol y.
 \label{eq:centroid2d}
 \end{aligned}
\end{equation}
These observables now follow from the same two dimensional mapping as the
physical flux magnification.

\begin{table}[t]
 \caption{Two dimensional weak lensing observables for an equatorial observer
 and a collinear source at $a_*=0.9$, $\gamma=0.1$, $\epsilon=0.01$ and
 $y_1=0.5$.  Although the images lie on $x_2=0$, their signed magnifications
 use the determinant of the full two dimensional Jacobian.  Separation and
 centroid are in units of $\theta_E$.}
 \label{tab:magnification}
 \centering\small
 \resizebox{\columnwidth}{!}{%
 \begin{tabular}{lrrrr}
 \toprule
Model & $\mu_-$ & $\mu_+$ & $\Delta\theta/\theta_E$ &
$\theta_{\rm c}/\theta_E$\\
 \midrule
Schwarzschild (1PN) & $-0.595017$ & 1.588316 & 2.090292 & 0.722116\\
$\lambda=0.5$ & $-0.611358$ & 1.604693 & 2.088288 & 0.718968\\
$\lambda=1$   & $-0.611298$ & 1.604753 & 2.087722 & 0.718966\\
$\lambda=1.5$ & $-0.607342$ & 1.601978 & 2.077436 & 0.719005\\
 \bottomrule
 \end{tabular}
 }
\end{table}

The full Jacobian changes the magnifications appreciably relative to the
scalar radial estimate because it includes the spin-dependent transverse
stretching.  The negative-parity image remains particularly sensitive, while
the centroid is nearly degenerate for this collinear benchmark.  This
illustrates why flux, separation and astrometric observables should be
analyzed jointly.

\begin{figure*}[t]
 \centering
 \includegraphics[width=1.\textwidth]{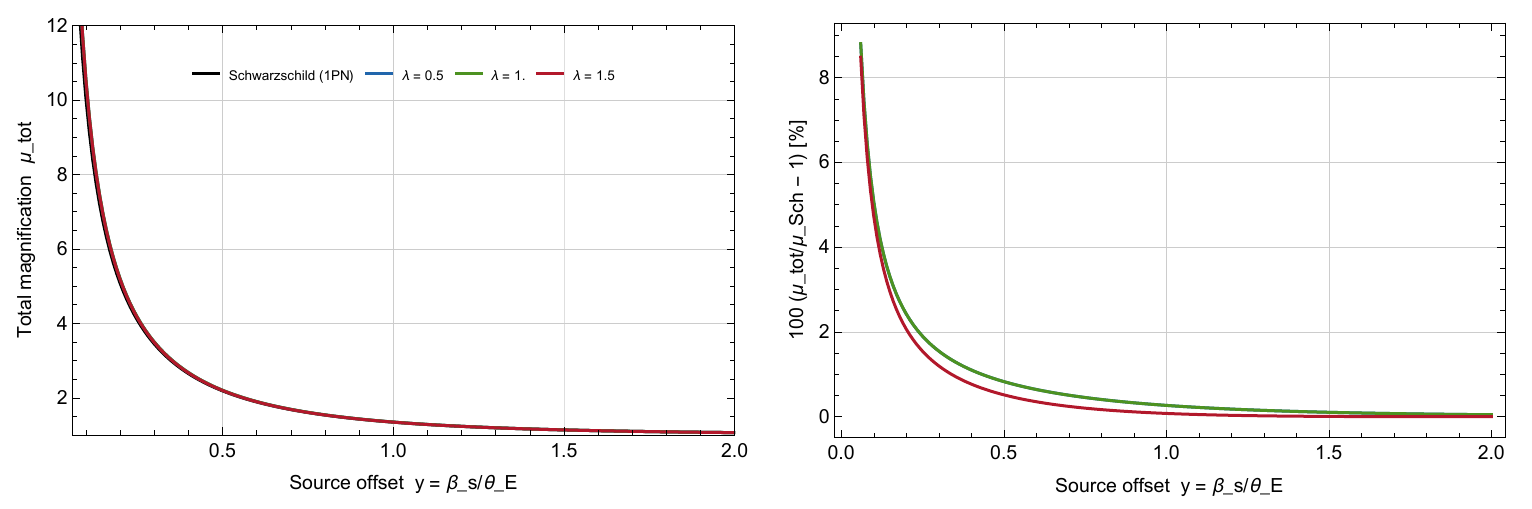}
 \caption{\emph{Physical weak lensing magnification obtained from the determinant of
 the two dimensional lens-map Jacobian and its fractional residual relative
 to the Schwarzschild result at the same post-Newtonian order.  The source is
 moved along the projected equatorial spin direction. The residual panel
 resolves differences compressed in $\mu_{\rm tot}$.}}
 \label{fig:mu}
\end{figure*}

The continuous source-position dependence of the unresolved flux is shown in Fig.
\ref{fig:mu}.  The first panel gives the absolute magnification, while the
second removes the dominant Schwarzschild contribution and thereby exposes
the much smaller KR-dependent residual.

The absolute curves in Fig. \ref{fig:mu} remain close because the point-mass term
dominates, especially near alignment.  The residual now contains both radial
and transverse responses of the rotating map and therefore represents a true
leading-order flux magnification.  Its
small magnitude also confirms that total magnification alone is a weak
discriminator, motivating combination with the signed magnifications,
separation and centroid in Table \ref{tab:magnification}.

\section{Conclusions and perspectives}

In this work, we investigated the causal, optical and effective-source
properties of the stationary and axisymmetric power law KR
geometry, generated from the asymptotically flat static solution.

We
reconstructed the exact inverse Einstein source associated with the rotating
metric and showed that it corresponds locally to a conserved anisotropic
type-I source with $p_r=-\rho$. We determined its principal stresses and
analyzed the corresponding energy conditions in the exterior region. For
positive $\Upsilon$, the energy density remained non-negative throughout the
range $0<\lambda\leq2$, while the radial null-energy condition was saturated.

The transverse dominant energy condition distinguished the underlying regions, considered in our analysis, being violated for $\lambda=0.5$,
saturated for $\lambda=1$ and satisfied outside the horizon for
$\lambda=1.5$ set up.

Further, we examined the curvature structure and
recovered the Kerr limit, as well as the original static power law
geometry, the Kerr-Newman-like geometry at $\lambda=1$ and the shifted mass
limit at $\lambda=2$.

For generic $\lambda$, we found that the singular
structure at $r=0$ extends beyond the usual Kerr ring. At the same time, the
inverse construction established the source required by the geometry without
identifying it with the stress tensor of a complete rotating KR
field configuration.

Afterwards, we characterized the horizon and stationary limits and derived
the extremality conditions for the rotating class of solutions. We quantified the size of the outer ergoregion through its angular dependent radial thickness and through both coordinate and proper volume measures.

In the slow rotation regime, the angular
thickness grew quadratically with the spin and reached its maximum on the
equatorial plane, whereas the proper volume displayed a leading linear
dependence on $|a|$ because of the near-horizon spatial geometry.

The exact
analysis showed that rotation provided the dominant control of the outer
ergoregion. The power law deformation produced a milder displacement of the
outer SLS, while it affected the inner horizon and inner
SLS more strongly. At rapid rotation, the dependence on
$\lambda$ also became visible in the equatorial thickness and in the
integrated ergoregion measures.

We worked out the separability of the Hamilton-Jacobi equation to derive the
spherical photon region and the corresponding critical impact parameters.
The equatorial photon regions display the expected frame-dragging
splitting, with the prograde orbit moving inward and the retrograde orbit
moving outward as the spin increased. Thus, we introduced signed
photon ergosurface gaps to quantify the position of these unstable photon
orbits relative to the outer SLS.

In particular, for positive
$\Upsilon$, the deformation shifted the penetration of the prograde photon
orbit into the ergoregion toward smaller spins. For $\gamma=0.1$, the critical
spin decreased systematically as $\lambda$ increased over the representative
regions, considered in our analysis. The retrograde photon orbit instead
remained outside the ergoregion throughout the investigated black hole
domain, showing an evident asymmetry between the co-rotating and
counter-rotating optical sectors.

From the unstable spherical photon region, we constructed the exact vacuum
critical curve for observers at arbitrary inclination and evaluated its
area equivalent diameter and horizontal displacement. Positive $\Upsilon$
reduced the enclosed critical curve area, while rotation generated the
dominant displacement of the image on the observer screen. For the
representative positive deformation, the area equivalent diameter decreased
with both spin and increasing $\lambda$, whereas its horizontal displacement
showed a substantially weaker dependence on the KR deformation.
We also obtained a small-deformation and slow rotation expression
for the fractional diameter shift and used the exact critical curve for the
high spin comparisons.

Within the asymptotically standard domain
$0<\lambda<2$, the comparison with M87$^\ast$ and Sgr~A$^\ast$ showed that
the present observational accuracy allowed sizable degeneracies among
$\lambda$, $\gamma$, spin and inclination.

We further analyzed the horizon surface gravity and the response of the
geometry under variations at fixed angular momentum. In this respect, direct comparison
between the surface-gravity temperature and a formal first law construction
shows a non-vanishing residual for generic $\lambda$. In this case, we found that the temperature defined from the
mass variation at fixed angular momentum and KR parameter did not
coincide with the temperature determined by the horizon surface gravity.
This difference showed that the standard first law identification could not be
assumed without specifying the appropriate entropy and conserved charges of
the rotating KR treatment.

We also derived the conditions governing Penrose
extraction and neutral wave superradiance and obtained the corresponding
maximum geometric estimates for rotational energy extraction.

Finally, we derived the rotating equatorial deflection angle and its
weak field expansion. The power law correction introduced a characteristic
impact-parameter dependence governed by $\lambda$, while rotation produced
the expected distinction between prograde and retrograde propagation.
Spin dependent KR terms entered at higher order and encoded the
coupling between frame dragging and the power law deformation.  We then
formulated weak lensing through a two dimensional thin lens map and computed
its full Jacobian, signed image magnifications, unresolved flux
magnification, image separation and astrometric centroid. The transverse
response induced by rotation produced corrections that were absent from a
scalar radial treatment. In the  configurations considered
here, the negative parity image shows the largest sensitivity, whereas the
total unresolved magnification remains  weak as evidence of the deformation itself.

In future works, we intend to construct an explicit rotating
KR two form compatible with the metric and the full nonminimally
coupled field equations. This development will extend our treatment through prime principles and will allow us to derive the conserved charges and Wald entropy, as well as a complete first law. We also intend to extend the
observational analysis through joint strong field and weak lensing scheme,
combining different outcomes. The main task of discriminating the KR parameters at both weak and strong gravity regimes will be central in our future developments, focusing also on the phenomenon of quasi-periodic oscillations \cite{qpo1,qpo2,qpo3,qpo4,qpo5} and within the accretion disks \cite{accretiondisk1,accretiondisk2,accretiondisk3,accretiondisk4}.

\vspace{0.5cm}

\begin{acknowledgments}
The authors acknowledge Konstantinos F. Dialektopoulos, Yergali Kurmanov, Andronikos Paliathanasis and Emmanuel N. Saridakis  for discussions on the topic of this work.
\end{acknowledgments}

\end{document}